# Cyber-Physical Control of Indoor Multi-vehicle Testbed for Cooperative Driving

Ali Bemani
Dept. of Electrical Engineering,
Mathematics and Science
University of Gävle
Gävle, Sweden
ali.bemani@hig.se

Niclas Björsell
Dept. of Electrical Engineering,
Mathematics and Science
University of Gävle
Gävle, Sweden
niclas.bjorsell@hig.se

*Abstract*—The system of connected vehicle to vehicle and vehicle to infrastructure can be considered as a wireless cyber-physical system of systems (Wireless CPSoS), which will be provided with the high ability of adaptive control on system of systems, cooperative scenarios to control of a Wireless CPSoS and adaptive wireless networked control system (WNCS). In this paper we present our multi-vehicle testbed based on the cyber-physical system that was designed for verification and validation of cooperative driving algorithm involving WNCS testing. Vehicles were developed as the physical prototype equipped with Raspberry-pi microprocessor and other sensing elements. This testbed consists of a fleet of 4 robot vehicles. An indoor positioning system (IPS) based on particle filter is purposed by using an inertial measurement unit (IMU) and iBeacon that is built upon Bluetooth Low Energy. Some typical cooperative driving scenarios can be implemented on this testbed under indoor laboratory. The method used to realize the objective statement was Model Predictive Control (MPC) with a state observer based on a Kalman Filter (KF). Because the wireless control systems can be severely affected by the imperfections of the wireless communication link. Our experimental testbed paves the way for testing and evaluating more intelligent cooperative driving scenario with the use new wireless technology and control system in the future.

## I. Introduction

System of systems (SoS) is a very broad area of research and the growing interest of that, made many challenges for system engineers [1]. SoS integrates several different systems into higher-level systems. This higher-level system pursues a goal that requires capabilities beyond the single capabilities of each system [2]. Systems of systems engineering is a concept of developing a process that is increasingly used for designing and realizing solutions to SoS challenges and persistent sociotechnical systems in areas such as industrial process, auto transportation, energy, global communication network, and many other SoS application domains [3].

Wireless networked control systems (WNCS) are composed of various sensor nodes, actuators and controllers that connect to each other over a wireless channel. The sensors measure the variation from physical plant and transfer their measurements to the controller and controllers take a decision to control the whole system by them. The actuators receive commands and influence on physical plant to keep the whole system stable and reliable [4], [5]. Cyber-Physical Systems (CPS) is a system with a compact interaction between computational environment and physical process to provide analysis techniques for the integrated whole and WNCS play a key role for control these systems [6]. Our idea is to combine WNCS, CPS and SoS to deal with the real-time control of the physical system of systems over the wireless networks. The main advantage of WNCS are the ease of installation in places where cabling is not easy, configurability and reduction of deployment and maintenance cost but also it has some imperfections like unreliable communication, packet loss and delays that these are critical for fast process. Also WNCS is a base technology for advanced control systems in automotive, industrial digitalization, critical control of a system of systems and Industry 4.0 [7], [8], [9].

The term Industry 4.0 refers to the subset of the fourth industrial revolution and is often known particularly as a framework for integration of CPS with real and virtual production, process, and services in the current industrial domain. In these systems, simulations are not only used for development. They are used inside the physical system involved in feedback to realize intelligent systems and grow up in environmental learning [10], [11]. Industry 4.0 introduces new design principles that the industry can organize itself accordingly, these basic principles are: increased interoperability between all components of smart factories through increase connectivity, virtualization of physical prototypes by connecting sensor data to the virtual models, decentralized control and decision making, real-time process to collect and analyze data for control in real-world system, modularity and increased service orientation [12], [13].

Industrial production has fundamentally changed with the principles of Industry 4.0 process and this trend has generated the need for introduction in academic guideline to support research and education. Therefore, it is very important to build proper experimental platform based on wireless cyber-physical system of systems to evaluate the real Industry 4.0.

In this work, we propose an efficient testbed that use multiple autonomous robots car to cooperatively driving on a train-vehicle. The testbed includes four vehicles with collision avoidance scenario and line tracker which communicate with a base station as a system of systems via wireless Ethernet. Also it is equipped with an indoor positioning system with accuracy up to cm-level and using IMU sensors with iBeacons received signal strength measurements based on particle filters and map matching algorithm.

Our goal in this paper is to develop an efficient testbed that allows us to test, implement and troubleshoot different control algorithms for a wireless CPSoS and provides the capabilities for students and researchers to study on WNCS imperfections and investigate how to overcome it for fast process. This testbed has also paved the ways for designing and implementing of a laboratory platform to evaluate difference between practical and theoretical understanding and verification of principles of Industry 4.0. In addition to the classical topics in self-control research, this laboratory environment provides new options and enable us to perform the following operations:

- Implementing a system of systems with a cloud platform, such that use the individual controller of each robot car will be employed to collect data and subsequently sending them to the overall controller for intelligent control and supervision of the systems.

- Design principles for Industry 4.0 based on the component, including above all the interoperability of manufacturing objects, virtualization of physical system, decentralization of the process, real-time capability and the modularity of manufacturing devices.

- Implementing a digital twins (DTs) as the next phase of CPS. Create a virtual model of the entire physical testbed in order to simulate and reflect their state through simulation analysis. A DTs is a one-to-one virtual replica of a system and it will be able to increase reliability, functionality and improve performance of the testbed by the data that comes from WNCS and IoT systems in this testbed. It will also be used to improve the overall manufacturing process.

- Design of a predictive maintenance algorithm based on data that is collected over time from the testbed to monitor the state of system and try to find patterns that can help to predict and ultimately prevent failures over the lifecycle of the testbed.

- Clarifying how exactly the reference architecture model for industry 4.0 (RAMI 4.0) working on testbed and provides a flexible service-oriented architecture framework combining services and data to understand the main aspects of Industry 4.0. RAMI 4.0 describes the core aspects of Industry 4.0 and defines Administration Shells as digital representations of Industry 4.0 components [14].

The remainder of the paper is organized as follows. Section II provides an overview of related experimental testbeds. Section III describe the entire system architecture and setup with the detailed description of the vehicle and environmental sensor suite. Then the whole cyber system architecture as a wireless networked control system is revealed in Section IV. Finally, Section V concludes the paper with highlighting the main results.

## II. RELATED WORK

There are several existing testbeds on CPS and Wireless CPSoS that provide a new area for research and development on industry 4.0. In this part, we focus on the existing testbeds from the structural viewpoint and execution techniques. Based on them the structure of all CPS testbeds can be divided into the three main components. Simulation-based, Hardware-based and Hybrid platform that in this work we focus more on Hybrid platform testbeds that are composed of physical components on the real-world and connected to the cyber components to communicate with each other in the virtual world. Several research group have recently developed based on this method. Such as in multi-agent vehicle system testbed [15], [16], automation robot testbed in industry [17], [18], [19], Unmanned Aerial Vehicle (UAV) system testbeds [20], [21].

In [15] a platform was designed that composed of multiple interacting intelligent agents with vehicle-to-vehicle (V2V) and vehicle-to-infrastructure (V2I) communication to evaluate the real connected vehicle testbed. The shortcoming for this testbed is the lack of a digital twin of this system for an evaluation in the virtual environment. Jiang and Zhou [16] proposed a system of connected vehicles that can be considered as a complicated CSP based on the hardware in loop simulation technology. They used ZigBee and Wi-Fi radio module to V2V and V2I communication and also used wireless network simulator to develop multi vehicle testbed that contains cross-layer cooperative communication simulation. This testbed was developed by the University of Waterloo and was designed to test and verify some typical cooperative scenarios and collaborative interaction under indoor laboratory environments. There are though also some deficiencies that have become evident, such as creating a virtual connection that connected virtual testbed to the physical testbed and low-accuracy of indoor positioning system.

Authors in [17] suggested a production machine testbed that consisted of a three-axis CNC engraving machine and controlled with Arduino. Additional sensors was added to this machine to measure the position of axes. A digital environment of this system was created by the use of information gathered from the sensors and was combined with a CAD model of the machine in a visual representation. This digital environment could be accessed from every location and all the information was stored in it. This testbed combined real and simulated information and is a real work of art toward realizing a testbed for digital, connected and adaptive production machine and production chain in industry 4.0. In the manufacturing system level [18], [19] respectively presented a digital twin of a machining process for a micro punching system and a self-acting barman robot. These manufacturing testbed, developed a cyber-physical manufacturing cloud, which combines cloud manufacturing and CPS for monitoring required signals in machine process and improve the quality and productivity of manufacturing.

Zhao et al [18] presented a context-aware autonomously controlling method of micro-dots punching machine tool via establishing the digital twin-driven cyber-physical system. This digital twin is proposed for supporting real-time synchronization of cyber and physical systems and improving the punching speed and positioning accuracy.

The testbed [20] is designed to abstract the control of physical system components to reduce the complexity of UAV oriented CPS experiments under realistic conditions with minimal cost. The positioning system in this testbed uses a single camera, mounted as a high point in the centre of the testbed and tracks the drones by tagging each drone with a distinct color tags. The UAV controller is responsible for moving the drone to a specific x, y coordination with the feedback that comes from the positioning system. A cyber-physical control of UAV at Sharif university of Technology [21] provides a testbed for testing a multivehicle cooperative control algorithm for control of UAV and the effect of time delay in the network communication. Based on the category of the architecture of CPS, this testbed is considered as a centralized structure if the algorithm run on the base station and is considered as a decentralized architecture if the algorithm runs on each UAV. At the simulation level of this testbed, the UAV is able to maintain a predefined altitude by using the fuzzy rules and adoptive control.

Finally, the advancements in the development of a testbed for smart manufacturing and model-based analytics will be accessible through apps provided from the cloud is surveyed in [22], this testbed offers a validation platform from the powertrain manufacturing for heavy vehicle (PMH) industry for cross-location development in industry 4.0 and smart production. This is an integrated hardware and software interface with a set of interconnected machine tools that data provided by the sensors and connectors and then will be stored in the digital twin to create live visualization of the production process. It was a great job for a testbed but based on the results this platform does not allow the user to implement the new algorithms in industry 4.0 and verify a new scenario for digital twins.

As an example for a wireless CPSoS platform, we need a testbed that consists of a system of systems with fast dynamics and integrates wireless communication and control methods together. This testbed allows us to test and implement different scenarios to evaluate wireless CPSoS regarding both network and control elements.

### III. TESTBED ARCHITECTURE

#### A. Testbed Description

A block diagram illustrating the sub-systems and interfaces in our testbed is shown in Fig. 1. This testbed consists of a cyberspace and physical space. The cyberspace of this system includes a motion capture system that tracks the miniature vehicles and provides pose measurements by use of the sensor's data. A Kalman Filter uses to provide vehicle's state that includes $[x, y, \omega]$. We implemented a decentralized control loop for the line tracking system to compute the steering angle set point. Also, we implemented a centralized control loop in cyberspace to compute the vehicle velocities according to their distance with the use of the LQR controller for controlling the distance between the leader car and the follower car in a train-vehicle cooperative control algorithm. An Indoor Positioning System based on Bluetooth beacons is used to provide real-time feedback on the position of the vehicle.

We use PiCar-s from SunFounder Company as a miniature vehicle to research for the physical prototype of hardware in this testbed.

The main component of the PiCar-s is a Raspberry PI. This miniature vehicle is also equipped with the ultrasonic obstacle avoidance module, the line follower module, and the inertial measurement unit to determine vehicle acceleration. An example of this miniature vehicle for realizing the hardware concept is depicted by (Fig. 2). As the miniature vehicles that are proposed in this testbed, they should be able to talk with the overall system controller through V2I communication. So, there are four major tasks for Raspberry PI perform, including DC gear motors drive, servo motor drive, sensors data processing, and Bluetooth and Wi-Fi communication.

The architecture of a digital twin-driven cyber-physical system for cooperative driving of multi vehicles is proposed as shown in Fig. 3. In this testbed, Raspberry PI used as a decentralized controller on each miniature vehicle to collect the ultrasonic and IR sensors data and forward them to the MATLAB as a centralized controller. Also, we use an IPS based on particle filters to provide real-time feedback on the vehicle's position that will be explained in detail in the next part. These data (position data, distance data, IR sensors data) are sent to the centralized controller over the wireless communication by the Raspberry PI, data is received by MATLAB and control algorithm is done with the use of Simulink environment and a reasonable physical model of the miniature vehicle. The state space of the system as two cars that follow each other is as follows. The state of each vehicle is its absolute position $p_i$ and velocity $v_i$, and its acceleration $a_i$ is the control input.

$$\begin{bmatrix} \dot{p_1}(t) - \dot{p_2}(t) \\ \dot{v_1}(t) \\ \dot{v_2}(t) \end{bmatrix} = \begin{bmatrix} 0 & 1 & -1 \\ 0 & 0 & 0 \\ 0 & 0 & 0 \end{bmatrix} * \begin{bmatrix} p_1(t) - p_2(t) \\ v_1(t) \\ v_2(t) \end{bmatrix} + \begin{bmatrix} 0 & 0 \\ 1 & 0 \\ 0 & 1 \end{bmatrix} * \begin{bmatrix} a_1(t) \\ a_2(t) \end{bmatrix}. \quad (1)$$

For this system, an LQR is designed with the full-state feedback, $Q = 1000I$ and $R = I$. To track the desired velocity, the desired state $\tilde{x}_{des}$ is introduced and used to implement the LQR law ($\tilde{u}(k) = \tilde{F}(\tilde{x}(k) - \tilde{x}_{des}(k))$). The velocity control values of each agent are sent to the miniature vehicles over wireless communication. The Raspberry PI applies the corresponding motors commands (DC gear motor and servo motor) with the pulse-width modulated signal. Also, the control algorithm on the centralized controller should be able to handle communication imperfections as will be discussed in detail in a subsequent section.

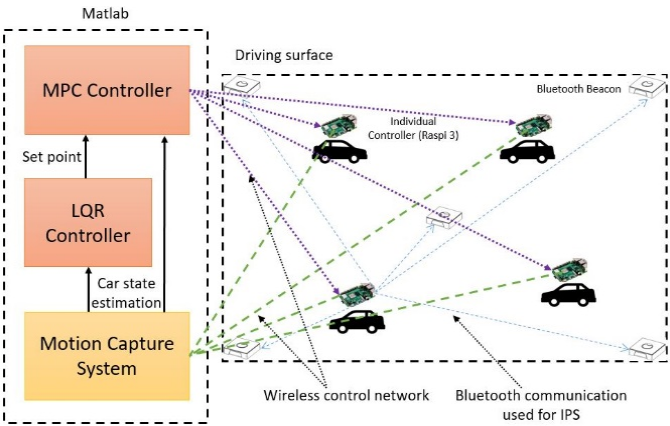

*Fig. 1. Testbed system architecture for showing sub-system and communications*

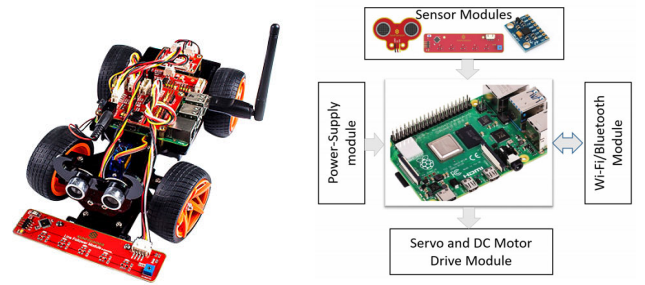

*Fig. 2. Vehicle realizing the hardware concept*

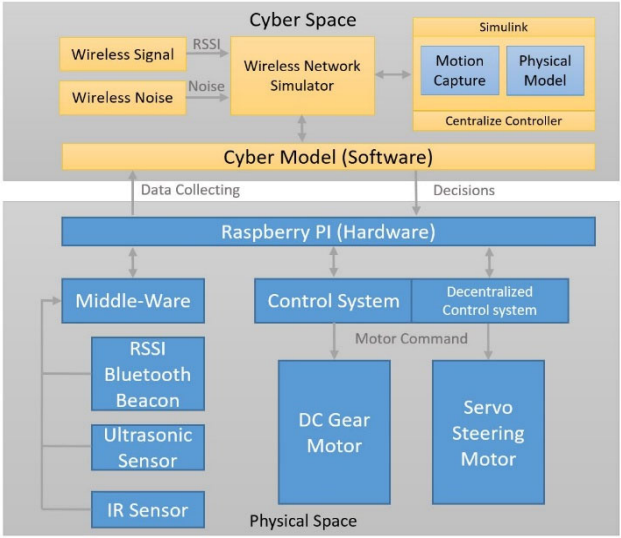

Fig. 3. Architecture of wireless cyber-physical system for the testbed

The proposed testbed provides us an open system of systems platform for evaluating cooperative driving algorithms from physical entities to cyber components. Train-vehicle is one of the cooperative experimental cases that was presented based on the proposed testbed. The following scenario is the four conditions that may happen to the distance between two cars and was implemented on the centralized controller as a train-vehicle:

- If the distance measured by the sensor is more than one meter, the miniature vehicle considers itself as a leader and start to move with 0.9 of the full speed.
- If the distance is less than one meter and bigger than 10 cm, the LQR controller will be activated and tries to keep the distance of 10 cm to the lead miniature car.
- If the distance is between 10 cm and 6 cm, the speed will follow a linear function of the distance.
- If the distance becomes less than 6 cm, the miniature vehicle stops.

A photograph of the testbed (3 m × 2 m) is shown in Fig. 4.

In this testbed, we propose a novel IPS using multiple sensing techniques, including IMU sensor and iBeacons that is built upon Bluetooth Low Energy. As we see in Fig. 5 we only deploy 5 iBeacons as landmarks with an exact position to cover our testbed. The cars will receive the received signal strength (RSS) value of the advertisement broadcast. The RSS value is used to calculate the distance between the sender (beacon) and receiver (cars). The RSS broadcast from an iBeacon can be formulated based on distance as:

$$RSS(d_j) = RSS(d_0) + 10n \log_{10}(\frac{d_j}{d_0}), \quad (2)$$

where $RSS(d_0)$ is the RSS value at the reference distance (1m), $RSS(d_j)$ is the RSS value that is received by $j^{th}$ car and n is the propagation path loss exponent. The distance $d_j$ between the car and the iBeacon can be calculated by:

$$d_j = d_0 * 10^{\frac{RSS(d_0)-RSS(d_j)}{10*n}}. \quad (3)$$

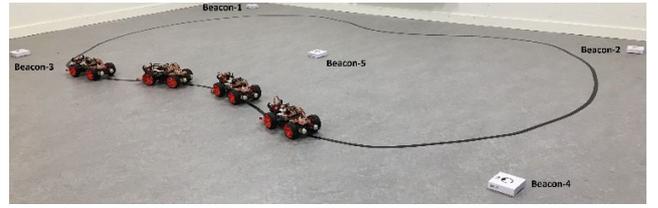

Fig. 4. Figure of the testbed

According to real data based on analyzing the possibility of error due to signal variation on the RSS from these iBeacons, we found that the RSS reading was not stable and reliable during the test. Therefore to overcome this inaccuracy, we used a Bayes filter that is able to filter out inaccuracies by estimating the unknown state with the help from current measurements and the previous known state according to the Hidden Markov Model and also with a map-matching algorithm.

To estimate the position, a particle filter is used. The main idea in particle filter is to represent the belief by a set of weighted random samples called particles. Also, we utilize an acceleration signal to estimate step detection in the car. A block diagram which extracts the position of the car using particle filter, map matching, acceleration data and distance to iBeacons has been developed in python, and is shown in Fig. 5. The accelerometer samples the acceleration data and sends it to the step detection component, the step detection algorithm is used to trigger the prediction model when a step has been detected. The RSS to distance component receives the RSS value from iBeacons ranging and converts them into the distance. The particle filter uses the step detection as a trigger for the prediction model and uses the RSS to distance component to update particles in the update model. The Map matching component is used to restrict the movement of particles in the prediction model. Its methodology is summarized as follows:

*1) Prediction Model:*

The state to track is $(x, y)$ in order to localize the cars. On start-up, the car has no information on its current position, so suppose we have an initial position for each car $(x_0, y_0)$ which is selected randomly and $N$ particles are uniformly distributed over the configuration space.

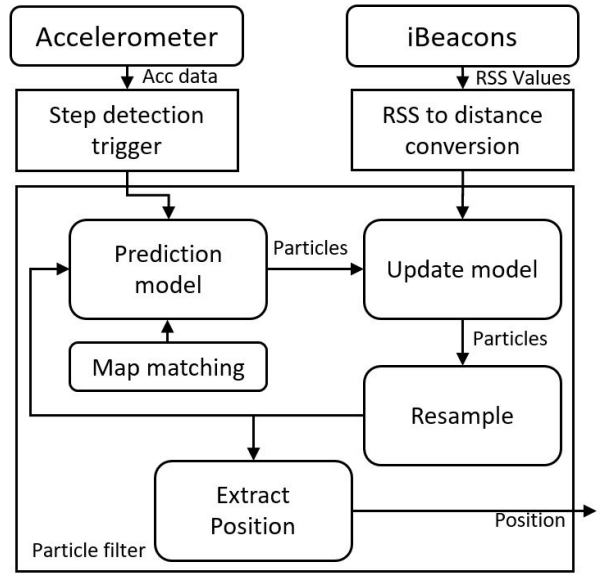

Fig. 5. Indoor positioning system algorithm

For each particle, we compute the probability that which one of these particles is closest to the car position. It assigns a weight $w_t^i$ for each particle proportional to this probability. The prediction model is responsible for moving the belief of the particle filter in the direction and with the length of the step. Then, the map matching component corrects parts of the belief that is unaccepted. After the new belief has been created, it is sent to the update model which includes the distance measurements from the iBeacons.

*2) Update model*

First, we assigned weight one to all of them then for every landmark that we have in our set of landmarks, we compute the distance between our particle and our landmark. Suppose the position estimation based on this algorithm at time step t is $(x_t^w)$. The distance between each particle and the landmarks is calculated as $\|x_k^L - x_t^i\|$ where $x_k^L$ is the position of the $k^{th}$ landmark. So we can define a distance $d_t^i$ for weight calculation of each particle. The weight of each particle is calculated as $w_t^i = w_{t-1}^i p(d_t|d_t^i)$, $i = 1, 2, \dots, N$ where $p(d_t|d_t^i)$ is a normal distribution.

$$p(d_t|d_t^i) = \frac{1}{\sqrt{2\pi\sigma^2}} e^{\frac{(d_t - d_t^i)^2}{2\sigma^2}}, \quad (4)$$

$(d_t)$ is the reported distance from the iBeacons and $d_t^i$ is the theoretical distance. So in the update model we try to compare these values, the smaller the difference is between the theoretical and the reported distance, a higher value is returned from the probability density function and then it is used to calculate the weight. Finally, in order to normalize the weights, each individual weight $w_t^i$ must be divided by the sum of all weights.

*3) Resample*

In the resample component, the particles with negligible weights are replaced by new particles in the proximity of the particles with higher weights. It causes that only the most likely particles survives to the next iteration of the particle filter.

*4) Extract Position*

For the final position estimation on each car, a weighted average is calculated from the new belief. The average is based on the weight and position of the particles.

*B. Testbed results*

The results of this testbed include two parts, simulation, and implementation. In the Simulation part to validate the LQR controller for controlling the distance between the two cars, the system is simulated with a square input signal as the acceleration. As shown in Fig. 6 the second car follows the first car by generating acceleration as its input model. The response of the LQR controller to the state-space model of two cars is shown in Fig. 7. As can be seen in the output of the model, with the movement of the first car, the second car also starts to move and keep the distance with it.

Fig. 8 is the cumulative distribution function of the positioning error for the IPS algorithm. As the Fig. 8 shows, the IPS algorithm performs very well in the whole testbed and most of the errors are under 10 cm. If 90% of the estimated positions are considered accepted, the position error in Fig. 8 shows that the IPS algorithm has a maximum error of 6 cm.

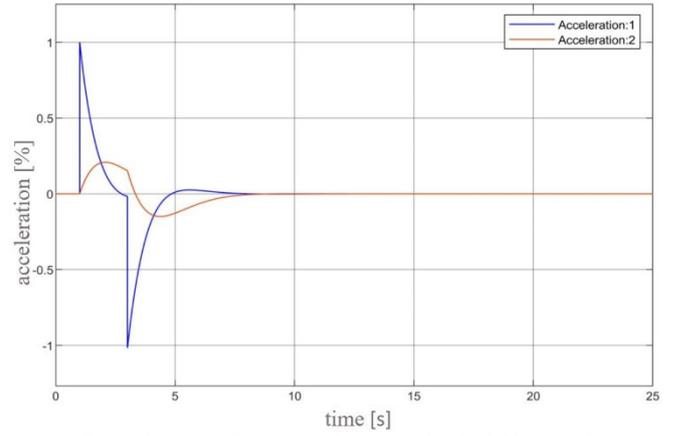

Fig. 6. Input acceleration to the system for the LQR control

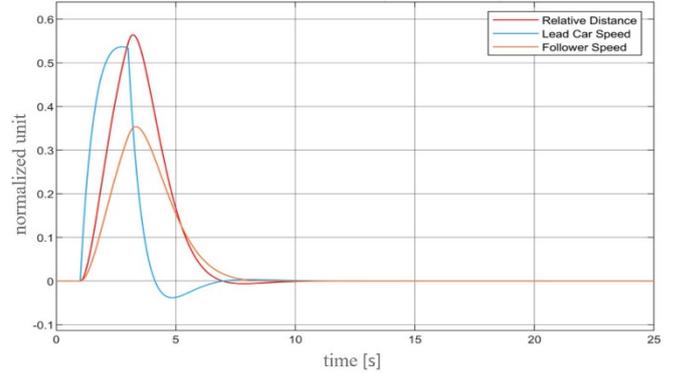

Fig. 7. Relative distance and velocity of LQR system

In the implementation part, the central controller in the cyberspace, where the model car adjusts its speed proportional to the relative distance of the car ahead with the LQR controller. The speed of each car is sent as a control signal over the wireless network and the follower car adjusts its distance to the car ahead. As shown in Fig. 9, except for the leader car, the rest of the cars follow the car ahead at a certain distance. The red dots indicate the position of each vehicle that is obtained by the IPS algorithm, the car's trajectory is shown in green and the blue dots indicate the position of each iBeacons. As can be seen from the car's trajectory, The IPS includes uncertainties because the RSSI measurements are affected by several parameters but these errors are less than 6 cm for the estimated positions.

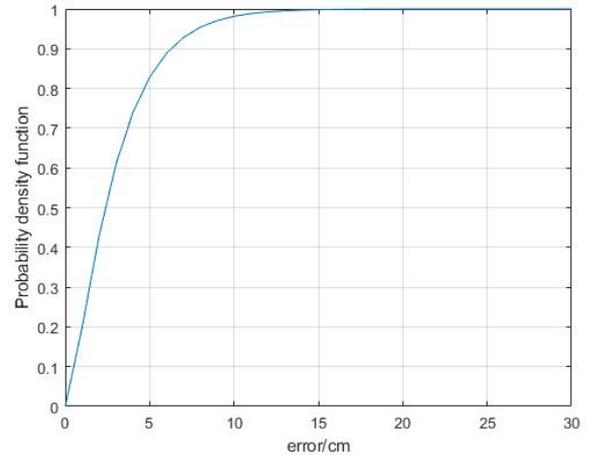

Fig. 8. The cumulative distribution function of positioning error

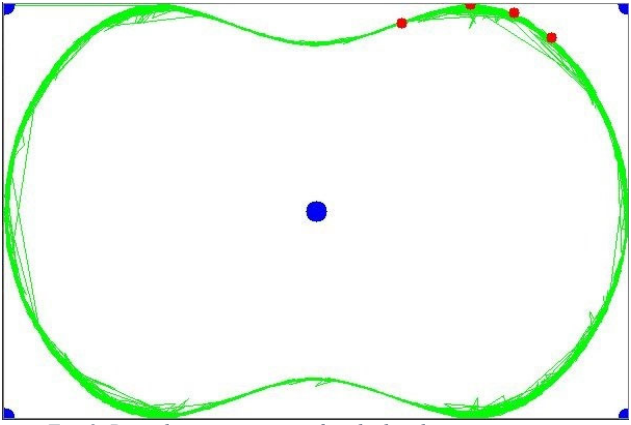
*Fig. 9. Digital representation of testbed with trajectories*

## IV. WIRELESS NETWORKED CONTROL DESIGN

As explained above, the miniature vehicles as physical systems, transmit the collected data from Raspberry PI to the centralized controller in the cyberspace through a single hop wireless network. Also, the controllers in the cyberspace tack decision and perform actions on miniature vehicles through a single hop wireless network. So, the sensor's data and the speed control commands need to be communicated over a wireless network, subject to delays and message losses.

For this testbed that the miniature vehicles use the same network for communication, another imperfection through in big scale is the limited bandwidth of the wireless channel. To mitigate these imperfections through wireless communications, the control design on the cyberspace integrates an observer based on a Kalman Filter with a model predictive controller. Then, in Raspberry PI, the control design integrates a buffer to store a sequence of speed control command computed by the overall controller [23]. The wireless networked control system architecture for this testbed is shown in Fig. 10. This system architecture is implemented on Simulink for each miniature vehicles. The cyber representation of this testbed is composed of multiple of this control loop that is connected to the overall controller in the cyberspace. (Fig. 11)

As a state observer for the closed-loop WNCS, we have implemented a Kalman Filter for full state estimation of each vehicle's physical model that includes $[x, y, \omega]$ with considering packet loss in the wireless network communication. In many wireless network control system, use a KF to achieve full-state estimation for optimal state feedback control under packet loss condition [24], [25].

The KF works in a two-step process: prediction step and update step. In the prediction step, the KF produces an estimate of the current state associated with the previous step, along with their uncertainties.

Then, this estimate is updated using a weighted average, with compares the current estimation with newly arrived sensing data. It is a recursive algorithm and can be run in real-time with using only the present input measurements, previously calculated state and the uncertainty of the sensors. In WNCS, when a sensing packet is dropped in the wireless network, the update step is skipped and the KF algorithm returning the same value as the prediction step. The KF is implemented with a similar theory in our WNCS design on the Simulink environment.

As a controller for the closed-loop WNCS, we have implemented an MPC that requires the iterative solution of Error optimal control problems on a finite prediction horizon. The main advantage of MPC is that it optimizes the system for the current sample and keeping the next samples in account [26]. We assume $T=100$ms as the sample time for the measurements of the sensors. At time $t$ the vehicle state is estimated by KF and a sequence of control commands for a relatively short time horizon in the future $[t, t + T]$ are computed by the MPC. We use this theory in Simulink and transmit this sequence to the buffer. This sequence is received by the buffer in decentralized vehicle control and feeds the physical model of the vehicle with a control input at each sample time. With MPC, if the wireless network drops a packet due to the imperfections, then the buffer feeds the physical model with the next available control input, which has been received in the last packet. As in Fig. 1, the KF takes the sensors data to produce estimated states $[x, y, \omega]$, the MPC takes these estimations, solved a discrete-time finite-horizon constrained LQR optimal control problem at each time step in the centralized controller and generates a sequence of predicted control set point.

The simulation typo is made up of a wireless network simulator and Simulink. TOSSIM is used as a wireless network simulator in a holistic simulation environment in cyberspace. The simulation-application protocol that is implemented in the cyberspace of the testbed is a command/event interface. We create a sample trace of received signal strengths and noise as inputs to the TOSSIM. The model in TOSSIM can replace a packet-level communication component for packet-level simulation.

To test and evaluate the control loop through the wireless control system in the cyberspace, we used harsh wireless conditions for noise trace in the model. With this idea, we can simulate the network with excessive packet drops. The Performance of this control loop is highly dependent on the accuracy of the KF and MPC. Based on the results, as shown in Fig. 12, when packet loss does occur, the control signal that is calculated by the MPC and stored in the buffer, applied to the system.

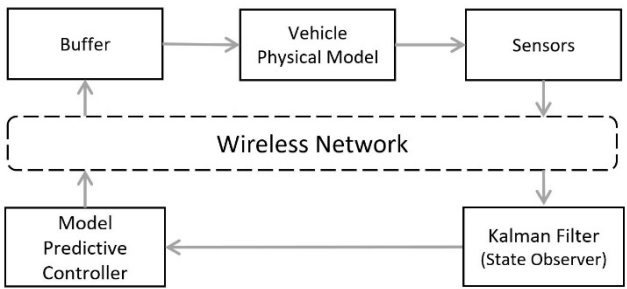
*Fig. 10. Wireless networked control system architecture for each miniature vehicle*

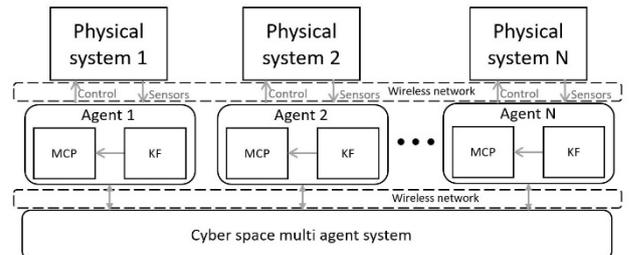
*Fig. 11. Multiple agents connected to the CPSoS*

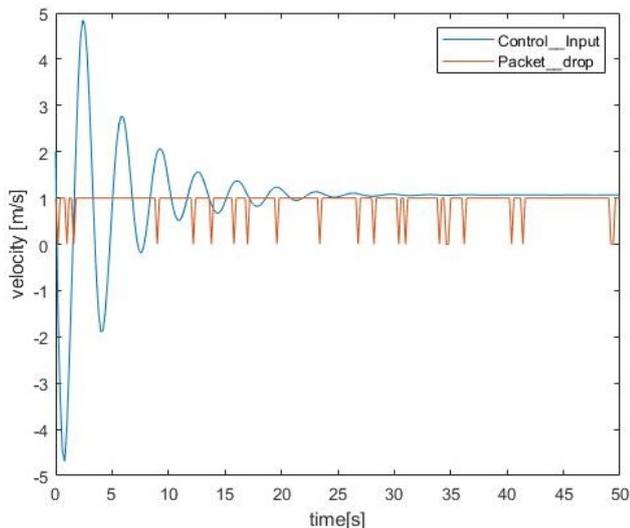

Fig. 12. Control input with packet loss compensation

The length of the buffer depends on the dynamics of the system under control and the characteristics of the wireless network. As can be seen, at intervals when packet loss occurs, this control signal is applied to the system without any effect of this failure.

## V. Conclusions

The goal of this work is to define a testbed with a wireless control network and a digital representation of a cyber-physical system of systems, which aims to focus mainly on wireless networked control between vehicle and infrastructure. The purposed testbed can be used to evaluate many different cooperative driving scenarios with the use of the digital representation of the system of systems in cyberspace. An accurate algorithm for RSSI indoor localization with the positioning error less than 6 cm was proposed in this paper. Then we implemented a control algorithm for the wireless networked control that has a different level of resilience to packet loss and is validated by the simulation results in the cyberspace.

## Acknowledgment

The research project is financed by the European Commission within the European Regional Development Fund, the Swedish Agency for Economic and Regional Growth, Region Gävleborg and the University of Gävle. We thank Amirhossein Hosseinzadeh for his work developing this testbed.